\documentclass[aps,preprint]{revtex4-2}

\usepackage{amssymb}
\usepackage{amsmath}
\usepackage{graphicx}
\usepackage{bm}
\usepackage{physics}
\usepackage{hyperref}
\usepackage{float}
\usepackage{subfig}  
\usepackage{subcaption}
\usepackage{dcolumn} 
\usepackage{caption}
\usepackage{natbib}  
\usepackage{xcolor}
\usepackage{geometry}        
\usepackage{float}           
\usepackage{overpic}         

\begin{document}

\title{How do higher-order interactions shape the energy landscape?}

\author{Zheng Wang}
 \affiliation{State Key Laboratory of Mechanics and Control for Aerospace Structures, College of Aerospace Engineering, Nanjing University of Aeronautics and Astronautics, Nanjing 210016, China}

 \author{Wenchang Qi}
 \affiliation{State Key Laboratory of Mechanics and Control for Aerospace Structures, College of Aerospace Engineering, Nanjing University of Aeronautics and Astronautics, Nanjing 210016, China}
 
\author{Jinjie Zhu}
\email{jinjiezhu@nuaa.edu.cn}
\affiliation{State Key Laboratory of Mechanics and Control for Aerospace Structures, College of Aerospace Engineering, Nanjing University of Aeronautics and Astronautics, Nanjing 210016, China}

\author{Xianbin Liu}
\email{xbliu@nuaa.edu.cn}
\affiliation{State Key Laboratory of Mechanics and Control for Aerospace Structures, College of Aerospace Engineering, Nanjing University of Aeronautics and Astronautics, Nanjing 210016, China}

\begin{abstract}

Understanding how higher-order interactions shape the energy landscape of coupled oscillator networks is crucial for characterizing complex synchronization phenomena. Here, we investigate a generalized Kuramoto model with triadic interactions, combining deterministic basin analysis, noise-induced transitions, and quantum annealing methods. We uncover a dual effect of higher-order interactions: they simultaneously expand basins for non-twisted states while contracting those of twisted states, yet modify potential well depths for both. As triadic coupling strengthens, higher-winding-number states and non-twisted states gain stability relative to synchronized states. The system exhibits remarkable stability asymmetry, where states with small basins can possess deep potential wells, making them highly resistant to noise-induced transitions once formed. These findings extend quasipotential theory to high-dimensional networked systems and offer new insights for controlling synchronization in complex systems.

\end{abstract}

\maketitle

\section{\label{sec:1}Introduction}

Understanding the collective behavior of coupled oscillators is fundamental to elucidating synchronization phenomena~\cite{10.1063/5.0176748,PhysRevLett.124.218301} across diverse systems, from neural circuits~\cite{dutta2019programmable,10.3389/fnsys.2022.908665} and cardiac tissue~\cite{doi:10.1126/sciadv.1701047,DOSSANTOS2004335} to power grids~\cite{doi:10.1126/sciadv.1500339,doi:10.1073/pnas.1212134110} and chemical reactions~\cite{Kuramoto1984}. The Kuramoto model~\cite{STROGATZ20001} has served as a paradigmatic framework for investigating these dynamics, offering analytical tractability while capturing essential features of synchronization transitions~\cite{skardal2020higher}. However, the complexity of real-world systems often extends beyond the pairwise interactions that characterize the classical model.

Recent studies have highlighted the critical role of higher-order interactions~\cite{battiston2022higher,battiston2021physics,doi:10.1137/21M1414024,PhysRevE.105.L042202}, where three or more units interact simultaneously in ways that cannot be decomposed into pairwise couplings. Such higher-order interactions have been identified in various contexts—from nonlinear dendritic processing in neural networks~\cite{doi:10.1073/pnas.80.9.2799} to cooperative binding in molecular systems~\cite{MAO1994532}. Incorporating these interactions into dynamical models has revealed novel phenomena, including abrupt synchronization transitions, cluster synchronization, and chimera states that cannot emerge through pairwise coupling alone~\cite{DOSSANTOS2004335,doi:10.1073/pnas.1212134110,skardal2020higher}.

External noise, manifested as random fluctuations in real-world systems, introduces non-deterministic behavior to oscillator dynamics, potentially enabling transitions across stability barriers that would be insurmountable in deterministic scenarios~\cite{HOLDER201710,PhysRevE.111.L012202}. While noise effects have been studied in standard Kuramoto models with pairwise interactions, their interplay with higher-order interactions remains largely unexplored. 

A powerful conceptual framework for understanding these complex dynamics is the energy landscape perspective~\cite{RevModPhys.70.223,wellens2003stochastic,PhysRevE.95.032317}, which visualizes system behavior as motion along a potential surface with multiple attractors (stable states) separated by barriers. Recent work by Zhang et al.~\cite{zhang2024deeper} demonstrated a counterintuitive phenomenon in Kuramoto oscillators with higher-order interactions: these interactions simultaneously increase the linear stability of synchronization patterns while shrinking their basins of attraction, creating what they termed ``deeper but smaller" basins. This finding raises fundamental questions about how the energy landscape is shaped by the combined effects of higher-order interactions and noise.

The present work addresses these questions by examining how higher-order interactions modify the energy landscape of coupled oscillator networks in the presence of noise. We focus on a generalized Kuramoto model with triadic interactions, investigating synchronization transitions, basin stability, and the relative depths of potential wells associated with different dynamical states. By elucidating the mechanisms through which higher-order interactions reshape the energy landscape, this study contributes to both the fundamental understanding of synchronization phenomena and practical applications in designing and controlling complex oscillatory systems.

\section{\label{sec:2}Model and Relative Basin Sizes}

To investigate how higher-order interactions shape the energy landscape of coupled oscillator systems, we consider a network of identical oscillators with higher-order coupling terms. Starting from phase reduction theory~\cite{Kuramoto1984}, we can derive the general form of dynamical equations incorporating higher-order interactions~\cite{leon2025theory}:
\begin{equation}
\label{eq:1}
\dot{\theta}_j = \omega_j + \varepsilon \sum_{k,l=1}^{N} A_{jkl}\Gamma_{jkl}(\Delta\theta_{kj}, \Delta\theta_{lj}),
\end{equation}
where $\theta_j$ represents the phase of oscillator $j$, and $\omega_j$ denotes the natural frequency of oscillator $j$. The adjacency tensor $A_{jkl}$ determines the topological structure of triadic coupling, where $A_{jkl} = 1$ if and only if there exists a triadic interaction among oscillators $j$, $k$, and $l$ with $j \neq k \neq l$, and $A_{jkl} = 0$ otherwise. The parameter $\varepsilon$ is a small perturbation parameter that quantifies the coupling strength. The function $\Gamma_{jkl}(\Delta\theta_{kj}, \Delta\theta_{lj})$ depends only on phase differences.  When both the pairwise and higher-order coupling $\Gamma$-functions are restricted to their first harmonic terms, we recover the extended Kuramoto model.

For analytical tractability, we focus on a ring network with coupling range $r$, where~\cite{zhang2024deeper}:
\begin{equation}
\label{eq:2}
\dot{\theta_i} = \frac{\sigma}{2r} \sum_{j=i-r}^{i+r} \sin(\theta_j - \theta_i) + \frac{\sigma_\Delta}{2r(2r-1)} \sum_{j=i-r}^{i+r} \sum_{k=i-r}^{i+r} \sin(\theta_j + \theta_k - 2\theta_i).
\end{equation}
The constraint $i \neq j \neq k$ ensures that each triadic coupling involves three distinct oscillators. The parameters $\sigma$ and $\sigma_\Delta$ control the pairwise and triadic coupling strengths, respectively, with normalization factors $2r$ and $2r(2r-1)$ accounting for the number of connections. For identical oscillators with uniform natural frequency $\omega$, we set $\omega = 0$ by transitioning to a rotating reference frame. In our numerical simulations, we adopt $n = 83$ oscillators and coupling range $r = 2$. The key findings remain qualitatively unchanged for other parameter choices. The selection of $n = 83$ follows the convention established in earlier studies~\cite{PhysRevLett.127.194101,10.1063/1.4986156}, while $r = 2$ represents the minimal coupling range that permits nontrivial simplicial complexes—a widely studied class of hypergraphs in network science literature, which we explore further in subsequent sections.

Previous studies have established that in ring networks with pairwise coupling, multiple twisted states can emerge as stable attractors under certain conditions~\cite{10.1063/1.2165594}. A $q$-twisted state is characterized by the phase configuration:
\begin{equation}
\theta_m^{(q)} = \frac{2\pi m q}{n} + C,
\end{equation}
where $m = 0, 1, 2, \ldots, n-1$ is the oscillator index, $q$ is the winding number representing the number of complete phase twists around the ring, and $C$ is an arbitrary constant representing the global phase shift. The case $q = 0$ corresponds to complete synchronization, while larger $|q|$ values indicate increasingly complex spatial phase patterns. Due to the rotational symmetry of the system, twisted states with winding numbers $q$ and $-q$ are equivalent and exhibit identical stability properties. The existence and stability of these twisted states depend on the network topology and coupling parameters.

To systematically distinguish between twisted and non-twisted states, we adopt the classification method from Ref.~\cite{zhang2024deeper} and quantify ordered dynamics using the parameter $P_{\text{twisted}}$:
\begin{equation}
P_{\text{twisted}} = \frac{n_{\text{twisted}}}{n},
\end{equation}
where $n_{\text{twisted}}$ is the number of ordered oscillators and $n = 83$ is the total number of oscillators. Following Ref.~\cite{zhang2024deeper}, we determine whether an individual oscillator is ordered by computing its local order parameter $O_j$:
\begin{equation}
O_j = \frac{1}{2r+1} \sum_{k=j-r}^{j+r} e^{i\theta_k}.
\end{equation}
An oscillator $j$ is classified as ordered if $|O_j| \geq 0.85$; otherwise, it is deemed disordered. Using this classification, twisted states correspond to $P_{\text{twisted}} = 1$ (all oscillators ordered), while non-twisted states exhibit $P_{\text{twisted}} < 1$. Specifically, chimera states are characterized by intermediate values $0 < P_{\text{twisted}} < 1$, where spatially localized regions of coherent and incoherent oscillators coexist, whereas completely incoherent states have $P_{\text{twisted}} = 0$.

\begin{figure}[htbp]
\centering
\captionsetup{justification=raggedright, singlelinecheck=false}
\includegraphics[width=0.9\textwidth]{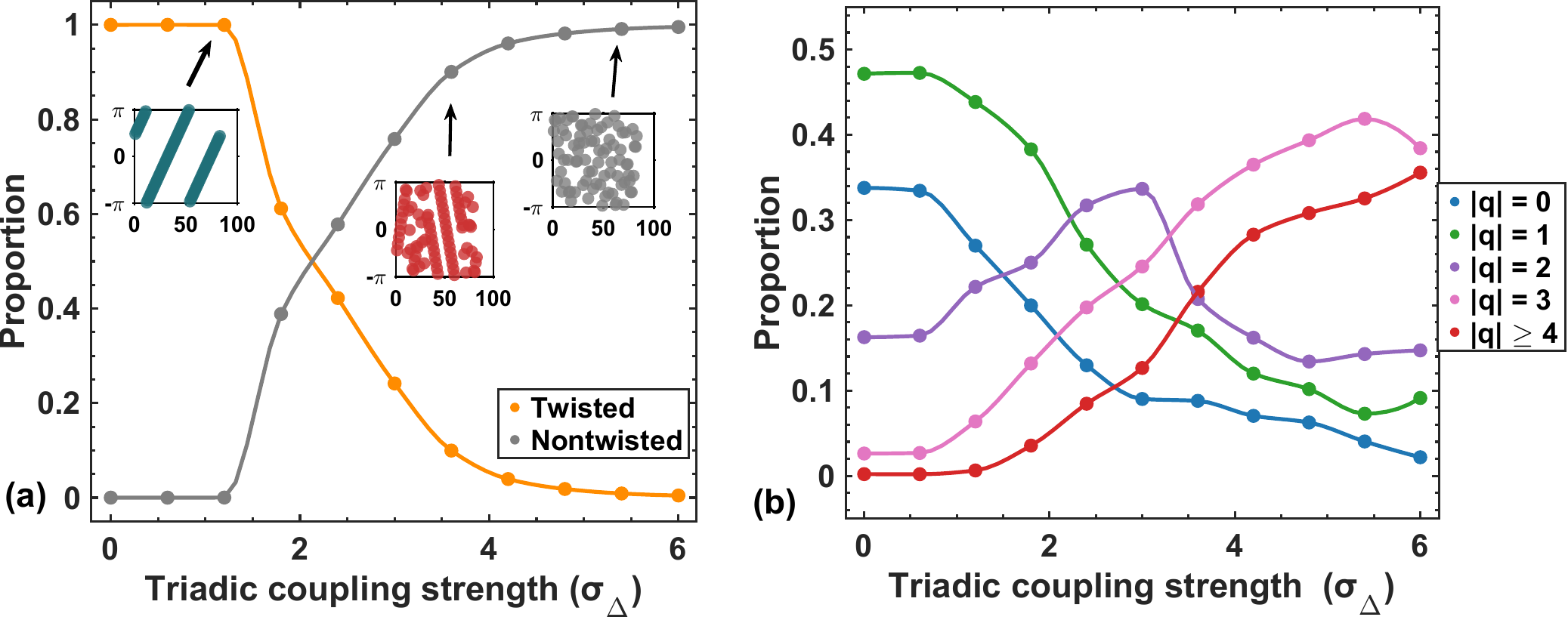}
\caption{Relative basin of attraction sizes as a function of triadic coupling strength $\sigma_{\Delta}$. (a) Proportion of state space occupied by twisted and non-twisted attractors. Insets show representative phase profiles with oscillator index on the horizontal axis and phase on the vertical axis, illustrating different dynamical states: twisted states (teal), chimera states (red), and incoherent states (gray). (b) Relative proportions of different twisted states within the twisted-state subspace, categorized by absolute winding number $|q|$. Numerical results based on $1.6 \times 10^5$ random initial conditions.}
\label{fig:1}
\end{figure}

Using this classification, we analyze how triadic coupling reshapes the basin structure of our oscillator network. Fig.~\ref{fig:1} illustrates how the relative sizes of these basins evolve with increasing $\sigma_{\Delta}$. When $\sigma_{\Delta}$ is small (approximately below 1.5), the entire state space is dominated by twisted states, with the fully synchronized state ($|q|=0$) and states with low winding numbers having the largest basins. As $\sigma_{\Delta}$ increases beyond this threshold, we observe a remarkable reorganization of the attraction structure—non-twisted states (primarily chimeras) emerge and their basins rapidly expand, eventually occupying nearly the entire state space. Simultaneously, the basins of attraction for twisted states significantly shrink. 

Further analysis of the twisted-state subspace reveals another intriguing pattern (Fig.~\ref{fig:1}b). In systems with only pairwise interactions ($\sigma_{\Delta}=0$), twisted states with lower winding numbers typically have larger basins of attraction. As $\sigma_{\Delta}$ increases, the relative proportion of low-winding-number states ($|q| = 0,1$) progressively decreases, while states with higher winding numbers gain prominence. 

\section{\label{sec:3}The Landscape of Basins of Attraction}

Given the existence of both twisted and non-twisted stable states, a natural question arises: how can we quantify the depth of potential wells associated with each type of attractor? To investigate this aspect of the energy landscape, we introduce Gaussian white noise into Eq.~\eqref{eq:2}, yielding the stochastic differential equation:

\begin{equation}
\label{eq:3}
\dot{\theta}_i = \frac{\sigma}{2r} \sum_{j=i-r}^{i+r} \sin(\theta_j - \theta_i) + \frac{\sigma_\Delta}{2r(2r-1)} \sum_{j=i-r}^{i+r} \sum_{k=i-r}^{i+r} \sin(\theta_j + \theta_k - 2\theta_i) + \sqrt{2}D\xi_i(t),
\end{equation}

\noindent where $D$ represents the noise intensity coefficient, and $\xi_i(t)$ are independent Gaussian white noise processes with zero mean and unit variance. 

To compare the relative depths of potential wells associated with different stable states, we can employ the Mean First Passage Time (MFPT)~\cite{doi:10.1137/0133024,PhysRevE.81.041119,gardiner2009stochastic}—a standard metric in stochastic dynamics that measures the average time required for a system to transition from one state to another under the influence of noise. The MFPT provides valuable insights into the stability landscape, with deeper potential wells corresponding to longer passage times.

\begin{figure}[htbp]
\centering
\captionsetup{justification=raggedright, singlelinecheck=false}
\includegraphics[width=1\textwidth]{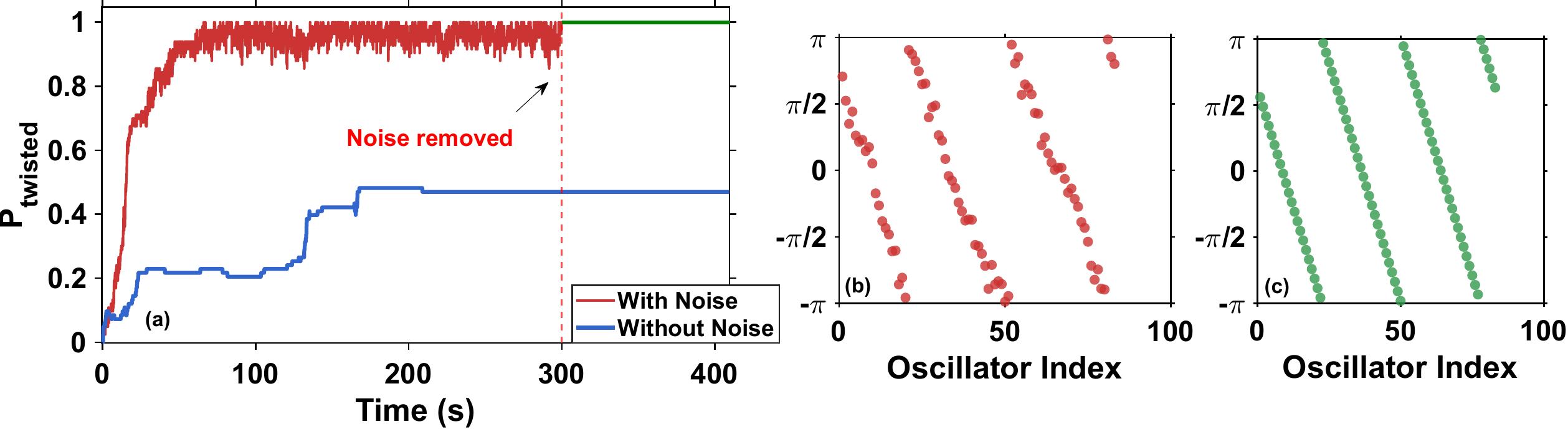}
\caption{Evolution of a system starting from a non-twisted initial state. (a) Temporal evolution of $P_{\text{twisted}}$ with noise (red curve) and without noise (blue curve). The blue curve confirms that the initial condition lies within the basin of attraction of a non-twisted state. In the case with noise, the green curve shows the system's behavior after noise removal at $t = 300$ s. (b) Phase configuration at the final time point under noisy conditions. (c) Phase configuration after 2 seconds of noise-free evolution, showing a clear twisted state pattern. Parameters: $\sigma_\Delta=3.5$, $D=0.4$.}
\label{fig:2}
\end{figure}

However, since higher-order networked systems do not conform to traditional multistable paradigms~\cite{doi:10.1142/S0218127408021233,PhysRevLett.52.9}, we must carefully define our initial conditions and escape criteria. First, we investigate measuring the MFPT from non-twisted states to twisted states. Our methodology is as follows: we first select a sample of random initial conditions that eventually stabilize into non-twisted states in the deterministic system. After introducing noise into the dynamics, we implement a noise-removal protocol at each time point—temporarily suspending the noise term for 2 seconds—to check if the system has transitioned to a twisted state. A successful transition is recorded if $P_{\text{twisted}}=1$ after this brief noise-free interval, as illustrated in FIG.~\ref{fig:2}.

The 2-second noise removal window serves a critical purpose: it allows the system to relax sufficiently close to a twisted state configuration if it has indeed crossed the basin boundary. We have verified that shorter or longer windows have minimal impact on quantitative results and do not affect the qualitative patterns observed in our analysis.

\begin{figure}[htbp]
\centering
\captionsetup{justification=raggedright, singlelinecheck=false}
\includegraphics[width=0.95\textwidth]{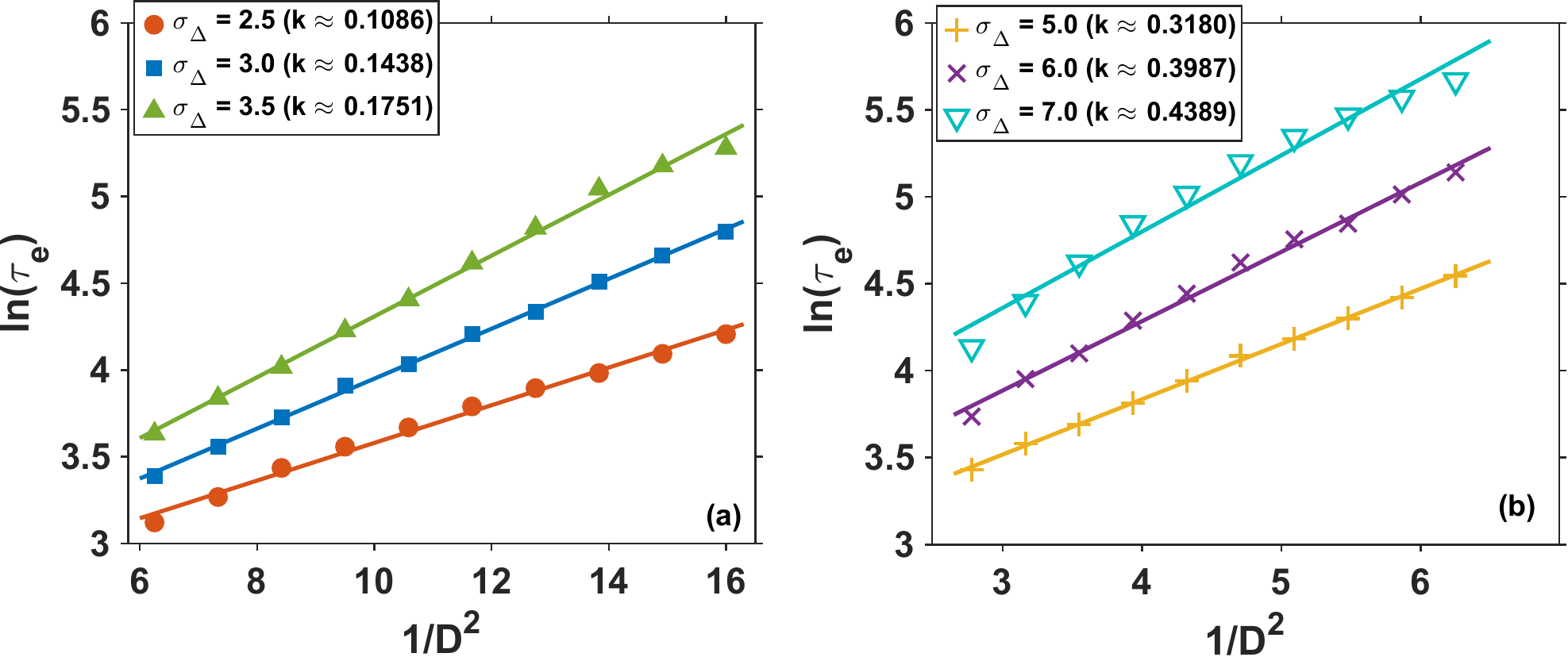}
\caption{Linear fitting of $\ln(\tau_e)$ versus $1/D^2$ for transitions from non-twisted to twisted states, demonstrating the exponential scaling relationship. (a) Results for moderate triadic coupling strengths ($\sigma_\Delta = 2.5, 3.0, 3.5$) with corresponding slope values ($k \approx 0.1086, 0.1438, 0.1751$). (b) Results for stronger triadic coupling strengths ($\sigma_\Delta = 5.0, 6.0, 7.0$) with corresponding slope values ($k \approx 0.3180, 0.3987, 0.4389$). Here, $k$ denotes the slope of the fitted line, which is proportional to the quasipotential barrier height in the small noise limit. Parameters: 2400 sample trajectories.}
\label{fig:3}
\end{figure}

To quantify the stability of attractors against noise-induced transitions, we analyze the First Passage Time (FPT) statistics. For our system, the FPT for a transition from a non-twisted state to a twisted state is defined as:
\begin{equation}
\begin{aligned}
\tau(\boldsymbol{\theta}_{\text{non-twisted}} \rightarrow \boldsymbol{\theta}_{\text{twisted}}) = \inf\{t \geq 0 \mid & \boldsymbol{\theta}(0) \in \text{non-twisted basin}, \\
& \boldsymbol{\theta}(t) \in \text{twisted basin}\},
\end{aligned}
\end{equation}
where $\boldsymbol{\theta}$ represents the full phase vector of the system. The MFPT is then calculated as the expected value of this random variable:
\begin{equation}
\tau_e = E[\tau],
\end{equation}
where $E[\cdot]$ denotes the averaging operator.

As shown in FIG.~\ref{fig:3}, the MFPT increases with increasing triadic coupling strength $\sigma_\Delta$. This indicates that higher-order interactions progressively deepen the potential wells of non-twisted states, making it increasingly difficult for the system to escape from these states to twisted states, thus elevating the MFPT.

The observed exponential relationship between MFPT and noise intensity suggests the existence of an effective barrier that the system must overcome during state transitions. This leads us to hypothesize that even in high-dimensional, non-gradient systems with higher-order interactions, a quasipotential~\cite{freidlin1998random,wellens2003stochastic,PhysRevE.48.931} framework might be applicable. The quasipotential concept, originally developed for low-dimensional systems, provides a scalar function that quantifies the minimum ``cost'' required for transitions between attractors in systems driven by small noise.

For a stochastic dynamical system, the quasipotential $\Psi(\boldsymbol{\theta}_A, \boldsymbol{\theta}_B)$ between states $\boldsymbol{\theta}_A$ and $\boldsymbol{\theta}_B$ can be defined as the minimum action required for the system to transition from $\boldsymbol{\theta}_A$ to $\boldsymbol{\theta}_B$:
\begin{equation}
\Psi(\boldsymbol{\theta}_A, \boldsymbol{\theta}_B) = \inf_{T>0} \inf_{\substack{\phi(0)=\boldsymbol{\theta}_A \\ \phi(T)=\boldsymbol{\theta}_B}} S_{0T}(\phi),
\end{equation}
where the action functional~\cite{freidlin1998random} $S_{0T}(\phi) = \frac{1}{2}\int_0^T|\dot{\phi}_t - b(\phi_t)|^2 dt$ represents the cost of deviating from the deterministic trajectory, with $b(\phi(t))$ denoting the deterministic vector field.

According to large deviation theory, the MFPT for transitions across the barrier separating attractors scales as~\cite{KRAMERS1940284}:
\begin{equation}
\tau_e \sim \exp\left(\frac{\Psi}{2D^2}\right),
\end{equation}
which aligns precisely with our empirical observations in FIG.~\ref{fig:3}. From this scaling relationship, we can extract the quasipotential barrier height:
\begin{equation}
\Psi \propto \frac{\ln(\tau_e)}{1/D^2},
\end{equation}

where the quasipotential barrier is directly reflected by the slope of the linear fit in the $\ln(\tau_e)$ versus $1/D^2$ plot. The slight increase in slopes with higher $\sigma_\Delta$ suggests that higher-order interactions modestly enhance the quasipotential barrier between non-twisted and twisted states.

\begin{figure}[htbp]
\centering
\captionsetup{justification=raggedright, singlelinecheck=false}
\includegraphics[width=0.95\textwidth]{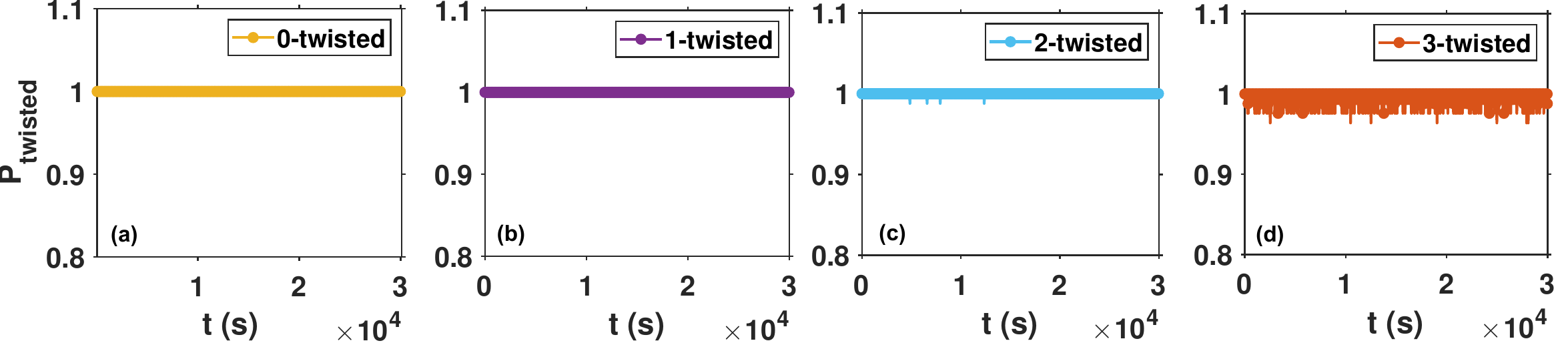}
\caption{Long-term stability of twisted states under noise. Temporal evolution of $P_{\text{twisted}}$ for different twisted states with winding numbers (a) $q=0$ (fully synchronized), (b) $q=1$, (c) $q=2$, and (d) $q=3$ under the influence of noise ($\sigma_\Delta=3.0$, $D=0.4$) for $3 \times 10^4$ seconds.}
\label{fig:4}
\end{figure}

This finding is particularly significant as it extends quasipotential theory to high-dimensional networked systems with higher-order interactions, providing a unifying framework to understand the seemingly contradictory effects of these interactions: they simultaneously deepen the potential wells of non-twisted states while expanding their basins of attraction.

\begin{figure}[htbp]
\centering
\captionsetup{justification=raggedright, singlelinecheck=false}
\includegraphics[width=0.5\textwidth]{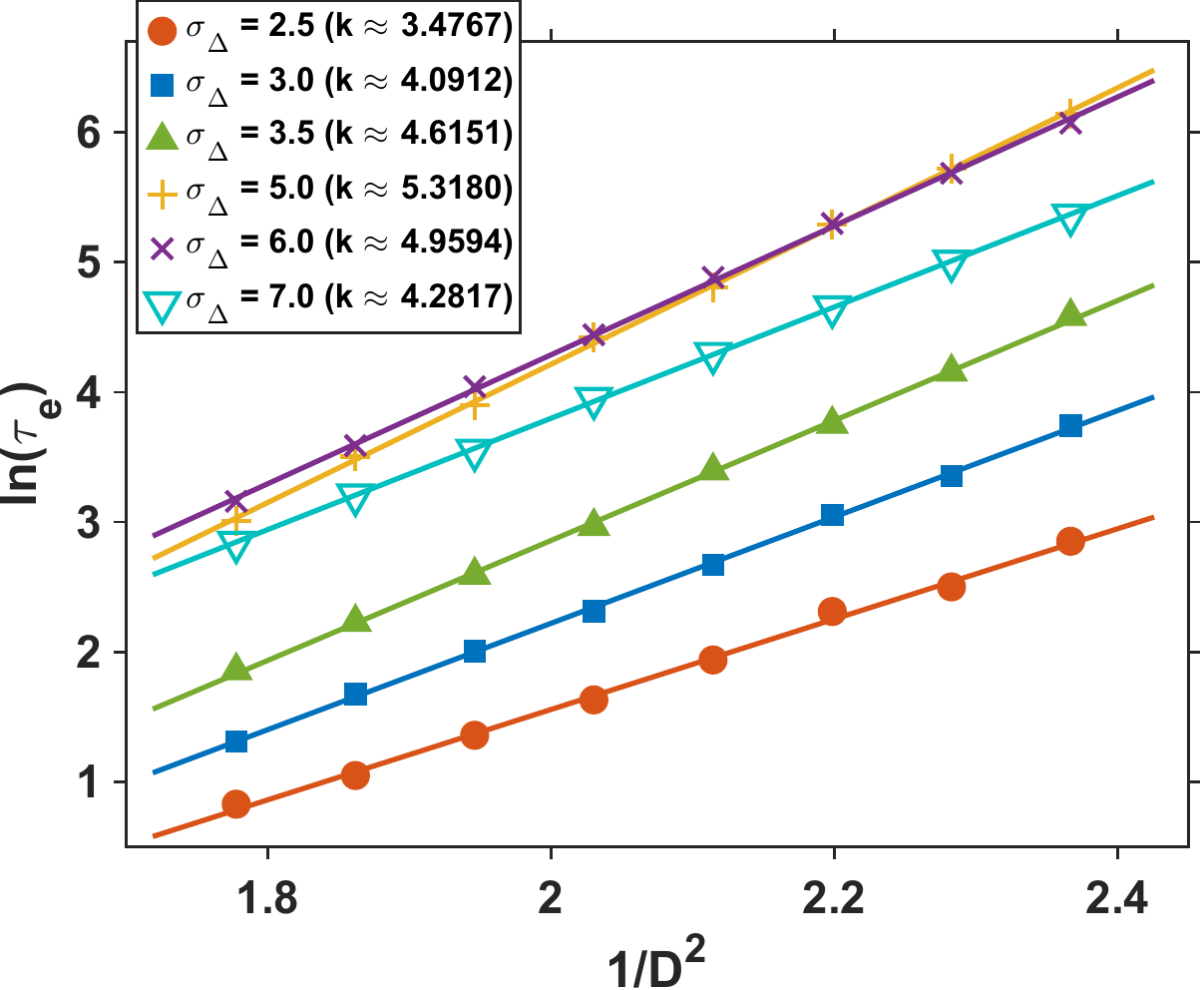}
\caption{Linear fitting of $\ln(\tau_e)$ versus $1/D^2$ for transitions from twisted to non-twisted state. Parameters: 2400 sample trajectories.}
\label{fig:5}
\end{figure}

Having examined transitions from non-twisted to twisted states, we now consider the reverse direction: transitions from twisted states to non-twisted states. Remarkably, we find that the system exhibits extreme resistance to such transitions. As shown in FIG.~\ref{fig:3}, when $\sigma_\Delta=3.0$ and $1/D^2=6.25$, the MFPT for transitions from non-twisted to twisted states is approximately 30 seconds. However, as demonstrated in FIG.~\ref{fig:4}, starting from twisted states with winding numbers $q=0,1,2,3$, the system shows no transitions to non-twisted states even after $3 \times 10^4$ seconds of simulation—a time scale three orders of magnitude longer. This striking asymmetry indicates that the potential wells associated with twisted states are substantially deeper than those of non-twisted states. 

To quantify this asymmetry more precisely, we calculated the MFPT for transitions from twisted to non-twisted states under various noise intensities. Our methodology involved initiating the system in twisted states, introducing noise, and then removing the noise for 5 seconds to check if the system had transitioned to a non-twisted state (defined by $P_{\text{twisted}} < 0.99$). The results, presented in FIG.~\ref{fig:5}, reveal exponentially longer transition times compared to the reverse process. Through additional verification not shown in the figures, we discovered a critical transition at triadic coupling strength approximately $\sigma_\Delta \approx 5.5$. Before this threshold, increasing triadic coupling progressively deepens the potential wells of twisted states, as evidenced by larger mean first passage times. However, beyond this threshold, this trend reverses—further increases in triadic coupling strength actually cause the potential wells of twisted states to become shallower, indicating a fundamental shift in the energy landscape's structure.

Furthermore, we observe an interesting pattern related to the winding number: as the winding number increases, the system shows greater fluctuations in $P_{\text{twisted}}$ due to noise (particularly visible in FIG.~\ref{fig:4}d), yet these fluctuations are insufficient to induce transitions to non-twisted states. The spikes visible in FIG.~\ref{fig:4}d are due to noise-induced perturbations temporarily affecting the order parameter, but the system quickly recovers without escaping the twisted state's basin of attraction. This profound stability asymmetry provides further evidence for the complex energy landscape shaped by higher-order interactions. While basins of attraction for twisted states shrink dramatically with increasing $\sigma_\Delta$, the depth of their potential wells remains substantial—creating a situation where twisted states have small basins of attraction yet exhibit high resistance to noise-induced transitions once formed.

Our analysis of mean first passage times has revealed striking asymmetries between twisted and non-twisted states, highlighting how triadic coupling reshapes the energy barriers between these distinct dynamical regimes. A critical question remains, however: within the family of twisted states, how does higher-order coupling influence the relative stability of configurations with different winding numbers?

The investigation of this hierarchical structure requires a methodological approach capable of systematically comparing potential well depths across the energy landscape. We implement a quantum annealing protocol~\cite{PhysRevE.58.5355,doi:10.1126/science.1057726,RevModPhys.90.015002} that leverages controlled quantum fluctuations to explore the complex stability landscape. This powerful technique enables precise characterization of relative stability among twisted states with varying topological complexity, providing insights into how higher-order interactions create preferential stability for certain winding number configurations.

Our annealing protocol involves systematically reducing the noise intensity in the stochastic system described by Eq.~\eqref{eq:3}. The procedure begins with a high noise intensity that enables the system to thoroughly explore the entire state space, effectively overcoming all potential barriers between different stable states. As the noise intensity gradually decreases, the system becomes progressively confined to deeper potential wells, with shallower wells becoming increasingly inaccessible. By analyzing the final distribution of states after the annealing process, we can infer the relative depths of their corresponding potential wells.

\begin{figure}[htbp]
\centering
\captionsetup{justification=raggedright, singlelinecheck=false}
\includegraphics[width=0.95\textwidth]{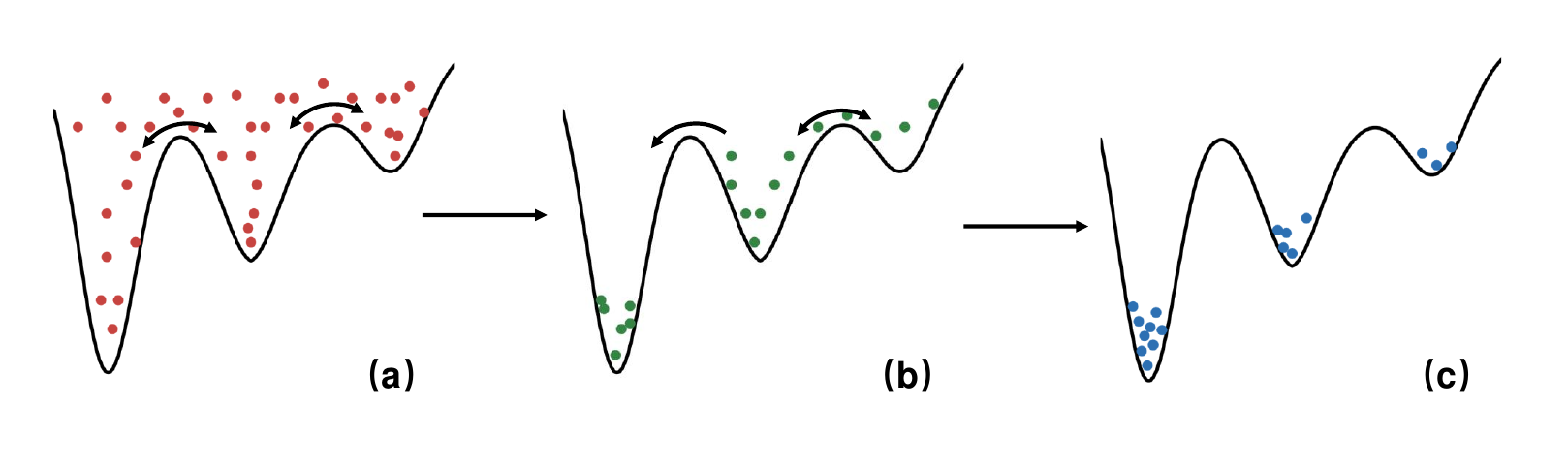}
\caption{Schematic illustration of the quantum annealing protocol for determining relative potential well depths. (a) High-noise phase: particles (red dots) exhibit ergodic behavior, exploring the entire energy landscape. (b) Intermediate phase: as noise decreases, particles begin accumulating in deeper wells while still traversing between shallower wells. (c) Low-noise phase: particles (blue dots) become trapped in potential wells according to their depths.}
\label{fig:6}
\end{figure}

The annealing schedule is defined as:
\begin{equation}
D(t) = D_0 e^{-t/\tau_A},
\end{equation}
where $D_0$ represents the initial noise intensity, $t$ is the time, and $\tau_A$ is the annealing time constant. We set $D_0 = 5.0$ to ensure comprehensive exploration of the state space and $\tau_A = 500$ to allow for gradual cooling.

FIG.~\ref{fig:6} illustrates the conceptual process of quantum annealing for determining relative potential well depths. The schematic depicts a multi-well potential landscape with three distinct stages of the annealing protocol:

(a) In the initial high-noise phase ($D = D_0$), particles (red dots) explore the entire energy landscape with sufficient energy to overcome all potential barriers. This thorough exploration ensures ergodicity— the system is not confined to any particular potential well and can access all possible states with equal probability.

(b) As the noise intensity gradually decreases (intermediate phase), particles become increasingly difficult to escape from the deepest well while still freely traversing between other shallower wells (indicated by the bidirectional arrow). During this phase, additional particles continuously migrate from shallower wells into the deepest one.

(c) In the final low-noise phase, virtually no transitions occur between wells. The final distribution of particles across different wells directly corresponds to the relative depths of these wells—deeper wells capture a larger proportion of particles.

This annealing approach provides a systematic method to quantify the relative stability of different attractors in our system. By comparing the final distribution of particles across different states, we can determine how higher-order interactions shape the hierarchical structure of the energy landscape. 

\begin{figure}[htbp]
\centering
\captionsetup{justification=raggedright, singlelinecheck=false}
\includegraphics[width=0.95\textwidth]{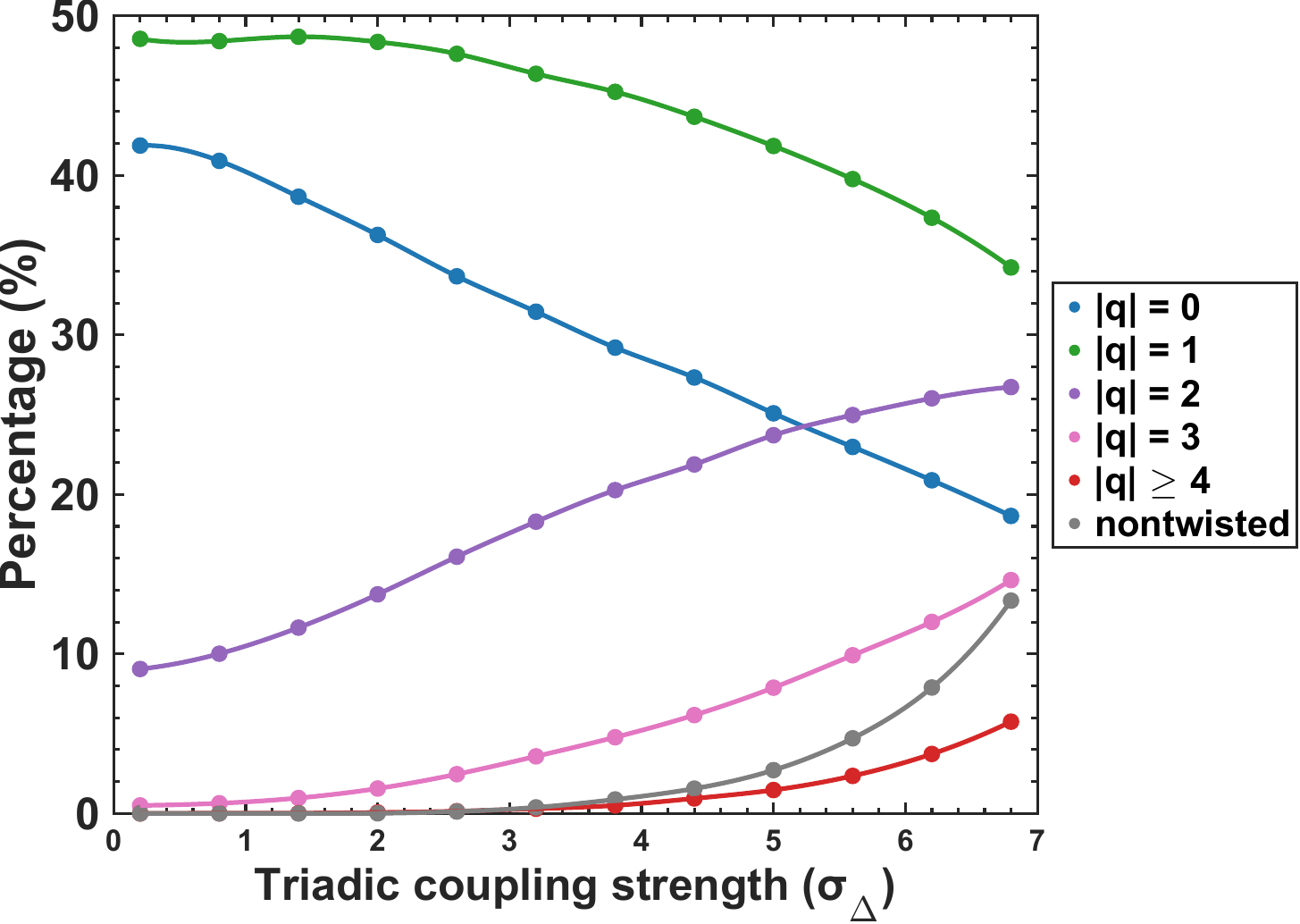}
\caption{Percentage distribution of different states as a function of triadic coupling strength $\sigma_\Delta$. Results were obtained through quantum annealing computations for 2000 s, followed by a 10 s noise-free evolution to determine the final state. Statistics based on 160,000 sample trajectories.}
\label{fig:7}
\end{figure}

FIG.~\ref{fig:7} shows the evolution of different state populations as a function of triadic coupling strength $\sigma_\Delta$, revealing clear trends in the relative stability of potential wells. As $\sigma_\Delta$ increases, we observe a systematic redistribution of state populations: the proportion of low-winding-number states ($|q|=0$ and $|q|=1$) steadily decreases, while the relative proportion of states with higher winding numbers ($|q|=2$ and $|q|=3$) shows an increasing trend. Notably, non-twisted states, which have negligibly small potential well depths at low coupling strengths, begin to develop significant stability when $\sigma_\Delta > 3$ and their relative proportion continues to grow as coupling increases. These results indicate that higher-order interactions progressively reduce the relative potential well depths of low-winding-number states while enhancing the stability of higher-winding-number and non-twisted states. 

These findings complement our previous observations and reveal an intriguing dual effect of higher-order interactions on the energy landscape. With increasing triadic coupling strength $\sigma_\Delta$, we observe both an expansion of the basin of attraction for non-twisted states and a contraction for twisted states (FIG.~\ref{fig:1}), while simultaneously increasing the potential well depths for non-twisted states (FIG.~\ref{fig:3}) and modifying those of twisted states (FIG.~\ref{fig:5}) in a non-monotonic manner, as well as altering the relative potential well depths among different winding numbers.

\section{\label{sec:4}Conclusion and Discussion}

In this study, we have systematically investigated how higher-order interactions reshape the energy landscape of coupled oscillator networks. By combining deterministic basin analysis, noise-induced transitions, and quantum annealing methods, we have uncovered a rich and nuanced picture of how triadic coupling fundamentally alters the system's multistability properties. Our analysis reveals several key findings that provide insights into the role of higher-order interactions in complex systems.

First, we demonstrated that increasing triadic coupling strength $\sigma_\Delta$ leads to a dramatic redistribution of basin sizes, with non-twisted states gaining dominance over twisted states. This topological transition is accompanied by a systematic change in the internal hierarchy among twisted states, where states with higher winding numbers constitute an increasing proportion of the twisted-state subspace, despite the overall reduction in twisted states as $\sigma_\Delta$ increases.

Second, our noise-based analysis revealed a striking asymmetry in the stability of different attractors. Despite their shrinking basins of attraction, twisted states exhibit remarkable resistance to noise-induced transitions, with potential wells that are substantially deeper than those of non-twisted states. This creates an intriguing scenario where states with the largest basins of attraction in the deterministic system (non-twisted states at high $\sigma_\Delta$) are also the most susceptible to noise-induced transitions.

Third, our quantum annealing approach provided a comprehensive view of the relative depths of potential wells across the state space. As $\sigma_\Delta$ increases, the system undergoes a systematic redistribution of potential well depths, with higher-winding-number states and non-twisted states gaining stability relative to the fully synchronized state. This finding challenges the conventional understanding that simpler topological configurations typically correspond to deeper potential wells.

Furthermore, we identified a dual effect of higher-order interactions on the energy landscape: they simultaneously modify both the sizes of basins of attraction and the depths of potential wells, often in seemingly contradictory ways. While previous work by Zhang et al.~\cite{zhang2024deeper} identified this phenomenon through linear stability analysis and termed it ``deeper but smaller'' basins, our approach using quasipotential theory provides a more comprehensive validation and refinement of this concept. Importantly, our findings demonstrate that the enhanced linear stability observed in prior studies indeed translates to genuinely deeper potential wells for non-twisted states, while twisted states exhibit a more complex, non-monotonic response to increasing triadic coupling strength, with potential well depths initially deepening but then becoming shallower beyond a critical coupling threshold. This distinction is noteworthy because linear stability analysis, while valuable, only characterizes local properties near equilibria and cannot fully capture such non-monotonic behaviors in the global energy landscape. Our systematic measurement of mean first passage times and their exponential scaling with noise intensity offers direct evidence of these nuanced barrier height changes, revealing stability properties that extend beyond the limitations of linear approximations. This reconciliation between local stability analysis and global energy landscape properties advances our fundamental understanding of how higher-order interactions reshape multistable systems.

These findings have several important implications. From a theoretical perspective, our work extends the applicability of concepts like quasipotential and energy landscape to high-dimensional networked systems with higher-order interactions. The demonstrated validity of the exponential scaling relationship between MFPT and noise intensity suggests that a scalar potential function can still effectively capture the essential stability properties of these complex systems, despite their non-gradient nature.

From a dynamical systems perspective, our results highlight the importance of considering both basin sizes and potential well depths when characterizing multistability. The observed decoupling between these two aspects—where states with small basins can nonetheless possess deep potential wells—challenges simplified views of attractor strength and suggests that a more nuanced approach is necessary for understanding stability in complex systems.

From an application's standpoint, the enhanced stability of higher-winding-number states under strong triadic coupling could be exploited in systems where maintaining topological diversity is desirable, such as in pattern formation, memory storage, or resilient network design. The ability of higher-order interactions to stabilize a wider range of spatial configurations might offer new avenues for controlling collective behavior in natural and engineered systems.

Several open questions remain for future investigation. First, how do these findings generalize to other types of higher-order interactions beyond triadic coupling? Second, can the observed stabilization of higher-winding-number states be leveraged for practical control strategies in real-world oscillator networks? Third, what are the implications of these energy landscape modifications for transient dynamics and information processing capabilities of neural or artificial networks with higher-order interactions?

In conclusion, our study provides a comprehensive characterization of how higher-order interactions shape the energy landscape of coupled oscillator systems, revealing a complex interplay between basin sizes and potential well depths that fundamentally alters the system's stability properties. These insights contribute to our understanding of collective behavior in complex systems and highlight the rich dynamical consequences of moving beyond pairwise interaction paradigms.

\section*{Acknowledgments}

JZ acknowledges the support from the National Natural Science Foundation of China (Grant No. 12202195). XL thanks the National Natural Science Foundation of China (Grant No. 12172167) for financial support.

\end{document}